\documentclass[12pt,a4paper,prd,amsmath,amssymb,showpacs,showkeys,nofootinbib]{revtex4-1}

\usepackage[utf8]{inputenc}
\usepackage[english]{babel}

\usepackage{bm}
\usepackage{mathrsfs}
\usepackage{mathtools}
\usepackage{geometry}
\usepackage{hyperref}

\newcommand{\bs}{\bm \sigma}
\newcommand{\bz}{\bm \zeta}
\newcommand{\bzf}{{\bz}^{(f)}}
\newcommand{\bn}{\bm \nu}
\newcommand{\bb}{\bm b}

\newcommand{\Sp}{\mathrm{Sp}}
\DeclarePairedDelimiter{\abs}{\lvert}{\rvert}
\DeclareMathOperator*{\RE}{Re}
\DeclareMathOperator*{\IM}{Im}
\DeclareMathOperator*{\Iunit}{i}
\DeclareMathOperator{\Exp}{e}
\renewcommand{\Re}{\RE}
\renewcommand{\Im}{\IM}
\renewcommand{\imath}{\Iunit}
\renewcommand{\exp}{\Exp}
\providecommand{\vect}[1]{\bm{#1}}

\allowdisplaybreaks[4]

\begin{document}

\title{On the polarization of fermion in an intermediate state}

\author{A.E. Kaloshin}
\email{kaloshin@physdep.isu.ru}
\affiliation{Physical Department, Irkutsk State University,
  K. Marx str. 1, 664003, Irkutsk, Russia}
\author{V.P. Lomov}
\email{lomov.vl@icc.ru}
\affiliation{Laboratory 1.2, Institute for System Dynamics and Control Theory of SB RAS,
  Lermontov str. 134, 664033, Irkutsk, Russia}
\affiliation{Irkutsk National Research Technical University,
  Lermontov str. 83, 664074, Irkutsk, Russia}

\begin{abstract}
  We show that calculation of a final fermion polarization (for a pure initial
  state) is equivalent to the problem of looking for complete polarization axis
  of bispinor. This gives the method for calculation of polarization applicable
  both for final and intermediate state fermions. We suggest to use fermion
  propagator (bare or dressed) in form of spectral representation, which gives
  the orthogonal off-shell energy projectors. This representation leads to
  covariant separation of particle and antiparticle contributions and gives a
  natural definition for polarization of intermediate state fermion. The most
  evident application is related with consistent description of $t$-quark
  polarization.
\end{abstract}

\maketitle


\section{Introduction}

It is well known how to calculate polarization of a final electron in framework
of quantum field theory, see for example \cite{Berestetskii:2012quantum}, \S65.
If we are interested in polarization of intermediate fermion, first of all we
need to give an accurate definition for this value. However, the concept of
polarization in intermediate state is used for a long time in particle
physics. One can recall the account of polarization in method of equivalent
photons \cite{Budnev:1974de}. Another example --- experimental and theoretical
activity concerning of polarization of $t$-quark produced in hadron collisions,
see experimental papers \cite{Aaltonen:2010nz, Abazov:2011gi,
  Chatrchyan:2013wua, Aad:2013ksa} and reviews
\cite{Schilling:2012dx,Bernreuther:2015wqa}. Note that in the last case the
naive definition for polarization is used in analogy with on-mass-shell
particle.

The first part of this paper concerns to technical aspects of calculation of a
fermion polarization. Namely, we discuss the problem of search for a complete
polarization axis \eqref{eq:27} for the final bispinor. It can be shown that the
found axis ($4$-vector $z$) coincides with polarization of final fermion
$s^{(f)}$ calculated by the standard way \eqref{eq:11}, \eqref{eq:14} for pure
initial state. Since the proposed method uses an amplitude but not its square,
it can be applied also for intermediate state. In this case one should simply
change an energy projector in a problem to the off-shell one.

The other part of this paper is devoted to application of proposed method for
intermediate fermion.  It is known that the on-mass-shell fermion density matrix
consists of two factors: the energy projector and spin density matrix
\cite{Bjorken:1964zz}. In order to define the fermion polarization in an
intermediate state, first of all one should introduce some off-shell energy
projectors. Here may exist different variants, but there is no some commonly
accepted definition. For example, in the known method of quasi-real electrons
\cite{Baier:1973ms, Baier:1980kx} the energy poles are accompanied by rather
artificial energy projectors which are not orthogonal to each other.

To give exact meaning for fermion polarization in an intermediate state, we
suggest to use the covariant form of propagator Eq. \eqref{eq:75}, which
contains the orthogonal off-shell projection operators. This view of free
propagator is a particular case of the spectral representation and can be
generalized with account of interactions. In Ref. \cite{Kaloshin:2011gy} the
spectral representation of fermion propagator in the theory with
$\mathsf{P}$-parity violation was constructed. Corresponding energy projection
operators (see \eqref{eq:43}) differ from the standard off-shell projectors and
also contain $\gamma^{5}$. Moreover, in such theory the standard spin projectors
do not commute with propagator and have to be modified as well
\cite{Kaloshin:2015rva}. The use of spectral representation for dressed
propagator gives natural definition for polarization of fermion in an
intermediate state.

The paper is organized as follows. Section~\ref{sec:nonrel} is devoted to
discussion of the problem of looking for complete polarization axis for
non-relativistic electron and its connection with calculation of a final
electron polarization as such. The equivalence of these two problems is shown
that allows to compute final electron polarization using the amplitude but not
its square.

In section~\ref{sec:rel} similar problem is discussed for bispinor. The
equivalence of these two problems is shown for relativistic electron that in
fact is consequence of similar property for two-dimensional spinors. As an
example the scattering of electron in external field is considered.

In section~\ref{sec:prop-spectral} the eigenvalue problem for inverse propagator
and corresponding spectral representation are discussed. This representation
leads to covariant separation of positive and negative energy poles which are
accompanied by the off-shell energy projectors orthogonal to each other. The
presence of $\gamma^{5}$ in vertex leads to modification (dressing) of both
standard energy and spin projectors.

Section~\ref{sec:intermed-fermion-polariz} is devoted to application of the
section~\ref{sec:rel} method for fermion in an intermediate state.  If to use
the spectral representation for propagator, the problem of search for a complete
polarization axis allows to find polarization vectors for particle and
antiparticle contributions. Let us note that the pure spin density matrices
correspond to these contributions.

In Conclusions we briefly discuss the obtained results. Some technical details
are grouped in Appendix.


\section{Scattering of non-relativistic electron}
\label{sec:nonrel}


\subsection{Polarization of final electron}
\label{sec:nonrel-m2}

Assume that electron from initial state characterized by twice the mean spin
vector $\bz$ is turned into final state characterized by vector $\bz'$
(polarization selected by a detector), then transition amplitude is
\begin{equation}
  f_{fi}=w^{\dag}(\bz')\, \hat{f} \,w(\bz).
 \label{eq1}
\end{equation}
Here $w(\bz)$ is eigenstate of operator $\bs \bz$:
\begin{equation}
  (\bs \bz)\, w(\bz)=w(\bz)
  \label{eq2}
\end{equation}
and similarly $w(\bz')$ is eigenstate of operator $\bs \bz'$. The operator
$\hat{f}$ in the general case is equal to $\hat f =A+\bs\vect{B}$, where $A$ and
$\vect{B}$ are complex parameters. For the sake of definiteness we choose
$\vect{B}=B\bn$ where $\bn$ is unit real vector such that
\begin{equation}
  \hat{f}=A+B\, (\bs\bn),\qquad \bn^{2}=1.
  \label{eq3}
\end{equation}

It is easy to calculate the $\bzf$, i.e. the polarization of final state as such
(see \S 140 in \cite{Landau:2013quantum}):
\begin{equation}
  \bzf=\frac{(\abs{A}^{2}-\abs{B}^{2})\,\bz+2\abs{B}^{2}\,\bn(\bn\bz)+
    2\,\Im(AB^{*})\, \bn\times\bz +2\,\Re(AB^{*})\, \bn}
  {\abs{A}^{2}+\abs{B}^{2}+2\,\Re(AB^{*}) (\bn\bz)}.
  \label{eq4}
\end{equation}
Note that Eqs. \eqref{eq1} and \eqref{eq2} are written for pure quantum states
for which $\abs{\bz}=\abs{\bz'}=1$. But the result \eqref{eq4} has exactly the
same form both for pure and mixed initial states, when $\abs{\bz}\leqslant 1$.

Few words to clarify the answer \eqref{eq4}. In order to calculate a final
electron polarization as such one needs the transition probability which is
proportional to square of matrix element \eqref{eq1}
\begin{equation}
  \abs[\big]{f_{fi}}^{2} = \Sp \Bigl( \frac{1+\bs \bz' }{2}\,\hat{f}\:
  \frac{1+\bs \bz}{2}\,\hat{f}^{\dag} \Bigr) = a + \bb \bz' =
  a \Big(1 + \frac{\bb}{a} \bz'\Big).
 \label{ME2}
\end{equation}
Here the terms containing detector polarization $\bz'$ and independent of it are
presented. The coefficients are calculated as:
\begin{equation}\label{coe}
  \begin{split}
    a   &= \frac{1}{2}\,\Sp
         \Bigl(\hat{f}\:\frac{1+\bs \bz}{2}\;\hat{f}^{\dag}\Bigr), \\
    \bb & = \frac{1}{2}\,\Sp
         \Bigl(\bs\,\hat{f}\:\frac{1+\bs \bz}{2}\,\hat{f}^{\dag}\Bigr).
  \end{split}
\end{equation}
The matrix element square \eqref{ME2} represents in fact the projection of spin
density matrix of scattered electron (which is characterized by some vector
$\bzf$) onto the detector spin density matrix $(1+\bs \bz')/2$. Thus, comparing
\eqref{ME2} with
\begin{equation}
  \Sp\Bigl(\frac{1+\bs \bz'}{2}\cdot
           \frac{1+ \bs \bzf}{2}\Bigr) =
  \frac{1}{2} (1 +  \bz' \bzf )
\end{equation}
one should identify the final electron polarization $\bzf$ as following:
\begin{equation}\label{eq:4}
  \bzf = \frac{\bb}{a}
\end{equation}
and we obtain the answer \eqref{eq4}.

Let us take a close look on the matrix
$\hat{f}\, \dfrac{1+\bs\bz}{2}\, \hat{f}^{\dag}$ which will be necessary
below. Using $\sigma$-matrix decomposition
\begin{equation}\label{eq:1}
  \begin{split}
    \hat{f}\, \frac{1+\bs\bz}{2}\, \hat{f}^{\dag} &= x_{0}+\bs\vect{x}=
    x_{0} \Big( 1+\bs \frac{\vect{x}}{x_{0}} \Big),\\
    x_{0}&=\frac{1}{2}\,
    \Sp \Big( \hat{f}\,\frac{1+\bs \bz}{2}\,\hat{f}^{\dag} \Big) = a,\\
    \vect{x} &= \frac{1}{2}\,
    \Sp \Big(  \bs\hat{f}\,\frac{1+\bs \bz}{2}\, \hat{f}^{\dag} \Big)=
    \vect{b},
  \end{split}
\end{equation}
one can see that the matrix is determined by the same parameters $a$ and $\bb$:
\begin{equation}\label{eq:2}
  \hat{f}\, \frac{1+\bs\bz}{2}\, \hat{f}^{\dag} = a+\bs \vect{b}=a(1+\bs \bzf).
\end{equation}


\subsection{The complete polarization axis of spinor}
\label{sec:nonrel-axis}

Let us consider an arbitrary two-dimensional spinor
\begin{equation}\label{eq:8}
  \chi=
  \begin{pmatrix}
    a\\
    b
  \end{pmatrix},\quad
  \abs{a}^{2}+\abs{b}^{2}=1.
\end{equation}
It is known (see \S 59 in \cite{Landau:2013quantum}) that for any spinor $\chi$
there exists an axis of complete polarization, i.e. there exists such unit
vector $\bz$ that
\begin{equation}
  \label{eq:3}
  (\bs\bz)\chi =\chi.
\end{equation}

In scattering process described by the amplitude \eqref{eq1} a new state appears
\begin{equation}
  w(\bz)\to \hat{f}\, w(\bz) = \hat{f}\,(1+\bs\bz)\,\chi,
\end{equation}
where $\chi$ is an arbitrary spinor. Let us consider the problem of search for
complete polarization axis $\vect{z}$ of spinor $\hat{f}\, w(\bz)$
\begin{equation}
  \big(\bs \vect{z} \big) \hat{f}\,(1+\bs\bz)\,\chi = \hat{f}
  \,(1+\bs\bz)\,\chi.
\end{equation}

It is convenient to rewrite this equation in an equivalent form
\begin{equation}\label{mat1}
  \frac{1 + \bs \vect{z}}{2}\cdot \hat{f}\, \frac{1+\bs\bz}{2} \chi =
  \hat{f}\, \frac{1+\bs\bz}{2} \chi .
\end{equation}

Recall that amplitude contains a spinor with definite polarization, so in Eq.
\eqref{mat1} vector $\bz$ is unit one (in contrast to \eqref{ME2}). However, if
to use a pure spin density matrix of initial state ($\bz^{2}=1$) within the spur
in \eqref{ME2}, then both problems, \eqref{ME2} and \eqref{mat1}, are equivalent
and two vectors coincide: $\bzf=\vect{z}$. Let us show this equivalence.

Suppose that we have solved the problem \eqref{mat1}, i.e. the vector $\vect{z}$
is known. Take Hermitian adjoint of the previous equation:
\begin{equation}\label{mat2}
  \chi^{\dag}\frac{1+\bs\bz}{2} \hat{f}^{\dag}\cdot
  \frac{1 + \bs \vect{z}}{2} = \chi^{\dag} \frac{1+\bs\bz}{2} \hat f^{\dag}.
\end{equation}
Multiplying Eqs. \eqref{mat1}, \eqref{mat2} by each other one obtains the
following matrix relation\footnote{The matrix
  $(1+\bs\bz)\chi \chi^{\dag} (1+\bs\bz)$ arisen after multiplication coincides
  up to a factor with $(1+\bs\bz)$. We come to \eqref{rel} after factor
  cancellation (recall that $\chi$ is an arbitrary spinor).}
\begin{equation}\label{rel}
  \frac{1 + \bs \vect{z}}{2}\cdot\Bigl( \hat{f}\,
  \frac{1+\bs\bz} {2}\, \hat{f}^{\dag}\Bigr)\cdot \frac{1 + \bs \vect{z}}{2} =
  \hat{f} \frac{1+\bs\bz}{2}\, \hat{f}^{\dag} \equiv X.
\end{equation}
Here the known  matrix $X$ \eqref{eq:2} has been appeared and previous
equation demonstrates its evident property:
\begin{equation}\label{eq:16}
  X\cdot(1-\bs \vect{z}) = (1-\bs\vect{z})\cdot X = 0.
\end{equation}
If to recall the matrix $X$ decomposition \eqref{eq:2}, we get two equations for
vector of final polarization $\bzf$ at the given $\vect{z}$
\begin{equation}\label{zsf}
  (1+\bs \bzf) \cdot(1-\bs \vect{z}) = (1-\bs\vect{z})(1+\bs \bzf) = 0.
\end{equation}

Writing down two relations \eqref{zsf} in detail:
\begin{equation}
  \begin{split}
    (1+\bs \bzf)\cdot(1-\bs \vect{z}) &=
    1-\bzf\vect{z}+\bs(\bzf-\vect{z})+\imath\bs(\bzf\times\vect{z})=0,\\
    (1-\bs\vect{z})\cdot(1+\bs \bzf) &=
    1-\bzf\vect{z}+\bs(\bzf-\vect{z})-\imath\bs(\bzf\times\vect{z})=0,
  \end{split}
\end{equation}
it is easy to see that the only solution is $\bzf=\vect{z}$.

On the other hand, if we know vector $\bzf$ then from \eqref{zsf} it follows that
$\vect{z}=\bzf$.

Thus we see that for scattering of non-relativistic electron these two problems:
calculation of the final electron polarization vector \eqref{ME2} (for pure
state) and looking for complete polarization axis \eqref{mat1} are
equivalent. So the final electron polarization can be calculated from scattering
amplitude instead of its square.


\section{Scattering of relativistic electron}
\label{sec:rel}

The scheme outlined above can be applied also for scattering of relativistic
electrons. Assume that electron from initial state characterized by momentum
$p_{1\mu}$ and polarization vector $s_{1\mu}$ is turned into final state
characterized by vectors $p_{2\mu}$ and $s_{2\mu}$ (polarization selected by a
detector), then transition amplitude is
\begin{equation}
  \label{eq:9}
  \mathscr{M}=\bar{u}_{2}(p_{2},s_{2})\Gamma u_{1}(p_{1},s_{1}).
\end{equation}
Here $u(p,s)$ is solution of Dirac equation
\begin{equation}
  \label{eq:10}
  (\hat{p}-m)u(p,s)=0,
\end{equation}
and matrix $\Gamma$ characterizes the transition process.

In order to find the final electron polarization it is necessary to
calculate the transition probability which is proportional to square of matrix
element \eqref{eq:9}
\begin{equation}\label{eq:11}
  \abs{\mathscr{M}}^{2}=\Sp \big(u_{2}\bar{u}_{2}\Gamma u_{1}\bar{u}_{1}
  \tilde{\Gamma}\big)=A+B_{\mu}s_{2}^{\mu}=A\Big(1+\frac{B_{\mu}}{A}s_{2}^{\mu}
  \Big),\quad
  \tilde{\Gamma}=\gamma^{0}\Gamma^{\dag}\gamma^{0}.
\end{equation}
Here we write down separately terms dependent on detector polarization $s_{2}$
and independent on it. Since $(s_{2} p_{2})=0$, only transverse part of vector
$B_{\mu}$ remains in \eqref{eq:11}
\begin{equation}
  B^{\perp}_{\mu} = \Big(g_{\mu\nu}-\frac{p_{2\mu} p_{2\nu}}{m^{2}}\Big)
  \cdot B^{\nu} .
\end{equation}

The matrix element square \eqref{eq:11} is in fact the projection of scattered
electron density matrix (its spin part is defined by vector $s^{(f)}_{\mu}$)
onto the detector density matrix $\rho'$. Thus comparison of \eqref{eq:11} with
\begin{equation}\label{eq:76}
  \Sp (\rho'\rho)=\Sp\Big(\frac{m+\hat{p}_{2}}{2m}\cdot
  \frac{1+\gamma^{5}\hat{s}_{2}}{2}\cdot
  \frac{1+\gamma^{5}\hat{s}^{(f)}}{2} \Big)=
  \frac{1}{2}\Big( 1-\big(s_{2}s^{(f)}\big) \Big)
\end{equation}
gives the final electron polarization $s^{(f)}_{\mu}$ as such:
\begin{equation}\label{eq:14}
  s^{(f)}_{\mu}=-\frac{B^{\perp}_{\mu}}{A}.
\end{equation}

Let us introduce short notations for final state projectors
\begin{equation}\label{eq:36}
  \begin{split}
    \Lambda^{\pm}_{2} & = \Lambda^{\pm}_{m}(n_{2})=
    \frac{1}{2}\big(1\pm \hat{n}_{2}\big), \quad
    \hat{n}_{2}=\frac{\hat{p}_{2}}{m},\quad
    n_{2}^{2}=1,\\
    \Sigma_{2} &= \Sigma_{0}(s_{2})=
    \frac{1}{2}\big(1+\gamma^{5}\hat{s}_{2}\big),\quad
    s_{2}^{2}=-1, \quad (s_{2} n_{2}) = 0 
  \end{split}
\end{equation}
and similarly for initial state. The matrix element square in these notations
is\footnote{Up to irrelevant numeric factor.}
\begin{equation}\label{eq:29}
  \abs{\mathscr{M}}^{2}=\Sp \big( \Lambda^{+}_{2}\Sigma_{2}
  \Gamma\Lambda^{+}_{1}\Sigma_{1}\tilde{\Gamma} \big)=
  \Sp \big( \Sigma_{2}X \big) =
  \Sp \Big( \frac{1}{2}\, (1+\gamma^{5}\hat{s}_{2}) X \Big).
\end{equation}
Here we have introduced necessary for further matrix (analogue of matrix
\eqref{eq:1} in the case of non-relativistic electron)
\begin{equation}\label{XXX}
  X=\Lambda^{+}_{2}\Gamma\Lambda^{+}_{1}\Sigma_{1}\tilde{\Gamma}
  \Lambda^{+}_{2} .
\end{equation}
The coefficients $A$ and $B_{\mu}$ in \eqref{eq:11} are calculated like
\begin{equation}\label{eq:12}
  A = \frac{1}{2}\, \Sp \big( X \big),\qquad
  B_{\mu} = \frac{1}{2}\, \Sp \big( \gamma^{5}\gamma_{\mu}X \big) ,
\end{equation}
and the orthogonality property $B_{\mu} p_{2}^{\mu} = 0$ is seen from it.

Let us find the decomposition of the matrix $X$ in $\gamma$-matrix basis. Simple
calculations show that all coefficients are easily expressed by means of
$p_{2}$, $A$, $B_{\mu}$ except of projection on $\sigma^{\mu\nu}$. The
decomposition has view
\begin{equation}\label{eq:37}
  X=\frac{A}{2}(1+\hat{n}_{2}) - \frac{1}{2}\gamma^{5}\hat{B}+
  \sigma^{\mu\nu} x_{\mu\nu},\quad
  \sigma^{\mu\nu}=\frac{1}{2}[\gamma^{\mu},\gamma^{\nu}].
\end{equation}
However, the coefficient $x_{\mu\nu}$ is related with parameters $A$, $B_{\mu}$
too. To see it, let us note that $X$ has the following properties
\begin{equation}\label{eq:39}
  (1-\hat{n}_{2})X=X(1-\hat{n}_{2})=0
\end{equation}
and the most general form of such matrix (see Appendix~\ref{sec:matrix-prob}) is
\begin{equation}\label{genX}
  X = (1+\hat{n}_{2})(x_{0} + \gamma^{5}\hat{x}) ,\qquad
  x_{\mu} n_{2}^{\mu} = 0 ,
\end{equation}
containing an arbitrary parameter $x_{0}$ and arbitrary 4-vector $x_{\mu}$
orthogonal to momentum.

Comparison of expression \eqref{eq:37} with the most general form \eqref{genX}
fixes  $\sigma^{\mu\nu}$ term. As a result the matrix $X$
\eqref{XXX} looks like
\begin{equation}\label{eq:40}
  X =\frac{1}{2} (1+\hat{n}_{2})\big( A - \gamma^{5}\hat{B} \big) =
  \frac{A}{2} (1+\hat{n}_{2}) \big( 1-\gamma^{5}\hat{B}/A \big).
\end{equation}
Recall that $s_{\mu}^{(f)}=-B_{\mu}/A$ is final electron polarization.


\subsection{The search for complete polarization axis of bispinor}
\label{sec:polaris-vector-matrix-problem}

After scattering described by the amplitude \eqref{eq:9} we have a new state
\begin{equation}\label{eq:26}
  u(p_{1},s_{1})\rightarrow \Lambda^{+}_{2}\Gamma u(p_{1},s_{1})=
  \Lambda^{+}_{2}\Gamma\Lambda^{+}_{1}\Sigma_{1}u(p_{1},s_{1}).
\end{equation}

Let us consider the problem of search for complete polarization axis $z_{\mu}$
of bispinor of scattered electron
\begin{equation}\label{eq:27}
  \gamma^{5}\hat{z}\cdot \Lambda^{+}_{2}\Gamma
  u_{1} = \Lambda^{+}_{2}\Gamma
  u_{1},\quad
  (zn_{2})=0 .
\end{equation}
We know in advance that this problem has a solution (see
Appendix~\ref{sec:bispinor-polar-axis-existence}). Let us rewrite the equation
in equivalent form
\begin{equation}\label{eq:28}
  \frac{1+\gamma^{5}\hat{z}}{2}\cdot \Lambda^{+}_{2}\Gamma
  u_{1} = \Lambda^{+}_{2}\Gamma
  u_{1}.
\end{equation}

Let us show that (as in the non-relativistic case) the complete polarization
axis $z$ coincides with $s^{(f)}$. Let us take Hermitian adjoint of previous
equation and multiply it by $\gamma^{0}$ from right
\begin{equation}\label{eq:30}
  \bar{u}_{1}\tilde{\Gamma}\Lambda^{+}_{2}\cdot
  \frac{1+\gamma^{5}\hat{z}}{2}=\bar{u}_{1}
  \tilde{\Gamma}\Lambda^{+}_{2}.
\end{equation}
Multiplying both equations by each other and substituting the density matrix of
initial electron $u_{1}\bar{u}_{1}$ one gets matrix relation
\begin{equation}\label{eq:31}
  \frac{1+\gamma^{5}\hat{z}}{2}\cdot \Big( \Lambda^{+}_{2}\Gamma
  \Lambda^{+}_{1}\Sigma_{1}\tilde{\Gamma}\Lambda^{+}_{2}
  \Big)\cdot \frac{1+\gamma^{5}\hat{z}}{2} =
  \Big( \Lambda^{+}_{2}\Gamma\Lambda^{+}_{1}\Sigma_{1}
  \tilde{\Gamma}\Lambda^{+}_{2} \Big) \equiv X .
\end{equation}
Note that here the known matrix $X$ \eqref{XXX}, \eqref{eq:40} has been appeared. This
relation can be transformed into equation connecting the complete polarization
axis $z_{\mu}$ and final electron polarization $s_{\mu}^{(f)}$.

As it follows from previous relation the matrix $X$ satisfies equations
\begin{equation}\label{eq:33}
  (1-\gamma^{5}\hat{z}) \cdot X = X \cdot (1-\gamma^{5}\hat{z}) = 0.
\end{equation}
If to use for $X$ the expression \eqref{eq:40} found above we obtain the
following two equations
\begin{equation}\label{eq:34}
  (1+ \hat{n}_{2}) (1-\gamma^{5} \hat{z}) (1+\gamma^{5} \hat{s}^{(f)}) = 
  (1+ \hat{n}_{2}) (1+\gamma^{5} \hat{s}^{(f)}) (1-\gamma^{5} \hat{z}) = 0 .
\end{equation}
Both expressions are ``under the observation'' of singular matrix
$(1+ \hat{n}_{2})$ but in the relation linking $z_{\mu}$ and $s_{\mu}^{(f)}$ it
does not play any role. Let us multiply spin matrices
\begin{equation}\label{eq:48}
  \begin{split}
    (1-\gamma^{5} \hat{z}) (1+\gamma^{5} \hat{s}^{(f)}) &= 1 + (z s^{(f)}) +
    \gamma^{5} \hat{s}^{(f)} - \gamma^{5} \hat{z} + \sigma^{\mu\nu} z_{\mu}
    s^{(f)}_{\nu} = 0, \\
    (1+\gamma^{5} \hat{s}^{(f)}) (1-\gamma^{5} \hat{z}) &= 1 + (zs^{(f)}) +
    \gamma^{5} \hat{s}^{(f)} - \gamma^{5} \hat{z} - \sigma^{\mu\nu} z_{\mu}
    s^{(f)}_{\nu} = 0.
  \end{split}
\end{equation}
It immediately follows that these two vectors coincide $s^{(f)}=z$.

So for relativistic electron there is also the equivalence of these two
problems: calculation of the scattered fermion polarization \eqref{eq:11},
\eqref{eq:14} (for pure initial state) and looking for complete polarization
axis \eqref{eq:28} of bispinor. The distinction of the problem of looking for
complete polarization axis is the use of amplitude instead of its square. This
makes possible to apply the same method for calculation of fermion polarization
both for final and intermediate states.


\subsection{Electron scattering in an external field}
\label{sec:scattering-in-ext-field}

By way of simple example let us consider single scattering of electron in an
external field. We will check the found above equivalence of two problems and
discuss some features.
\begin{center}
  \includegraphics{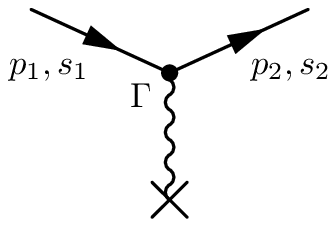}
\end{center}
Matrix element has form \eqref{eq:9} in which vertex factor $\Gamma$ contains
Fourier transform of the external field and corresponding $\gamma$-matrix.

The problem of looking for axis \eqref{eq:28} can be rewritten in equivalent
view
\begin{equation}\label{eq:49}
  \frac{1+\gamma^{5}\hat{z}}{2}\cdot \Lambda^{+}_{2}\Gamma\Lambda^{+}_{1}
  \Sigma_{1} \chi=\Lambda^{+}_{2}\Gamma\Lambda^{+}_{1}\Sigma_{1} \chi ,
\end{equation}
where $\chi$ is an arbitrary bispinor. Since this is an inverse problem
(reconstruction of an operator), $z^{\mu}$ generally speaking can depend on
bispinor $\chi$. However, the found equivalence $s^{(f)}=z$ and expression
\eqref{eq:14} for $s^{(f)}$ tell that vector $z$ does not depend on bispinor
$\chi$. It means that \eqref{eq:49} can be rewritten as a matrix problem
\begin{equation}\label{eq:59}
  \frac{1+\gamma^{5}\hat{z}}{2}\cdot \Lambda^{+}_{2}\Gamma\Lambda^{+}_{1}
  \Sigma_{1}=\Lambda^{+}_{2}\Gamma\Lambda^{+}_{1}\Sigma_{1}.
\end{equation}
The matrix formulation of the problem gives rather convenient method to find
vector $z$: one should decompose \eqref{eq:59} in $\gamma$-matrix basis and seek
for vector $z_{\mu}$ as an expansion over available vectors\footnote{For $S$ and
  $P$ vertices these are vectors $n_{1\mu}$, $n_{2\mu}$, $s_{1\mu}$ and
  expansion looks as
  \begin{equation*}
    z_{\mu}=z_{1}n_{1\mu}+z_{2}n_{2\mu}
           +z_{3}s_{1\mu} +
    z_{4}\epsilon_{\mu\nu\lambda\sigma}
    n_{1}^{\nu}n_{2}^{\lambda} s_{1}^{\sigma},
  \end{equation*}
  for $V$ and $A$ vertices there exists also vector $A_{\mu}$ and expansion is:
  \begin{equation*}
    z_{\mu}=z_{1}n_{1\mu}+z_{2}n_{2\mu}+z_{3}s_{1\mu}+z_{4}A_{\mu}.
  \end{equation*}
  Case of linear dependence of vectors needs special consideration.}%
. We checked that for external fields of different kinds ($S$, $P$, $V$, $A$)
the solution of the problem \eqref{eq:59} coincides with final fermion
polarization calculated according to \eqref{eq:14} (for definite polarization of
initial fermion, $s_{1}^{2}=-1$).

We will present here solutions of the problem \eqref{eq:59} for scalar and
vector vertices
\begin{itemize}
\item $\Gamma=1$
  \begin{equation}
    \label{eq:18}
    z_{\mu}=s_{1\mu}-a_{1}p_{1\mu}-a_{2}p_{2\mu},\quad
    a_{1}=a_{2}=\frac{(p_{2}s_{1})}{(p_{1}p_{2})+m^{2}}.
  \end{equation}
\item $\Gamma=\gamma^{\mu}A_{\mu}(q)$
  \begin{equation}
    \label{eq:19}
    \begin{split}
      z_{\mu} &= s_{1\mu}-a_{1}p_{1\mu}-a_{2}p_{2\mu}-a_{3}A_{\mu},\\
      a_{1} &= -a_{2}=\frac{(p_{2}s_{1})(AA)-2(p_{2}A)(s_{1}A)}{D},\\
      a_{3} &=2
      \frac{(p_{1}p_{2})(s_{1}A)-(p_{1}A)(p_{2}s_{1})-(s_{1}A)m^{2}}{D},\\
      D &= (p_{1}p_{2})(AA)-2(p_{1}A)(p_{2}A)-(AA)m^{2}.
    \end{split}
  \end{equation}
\end{itemize}

Let us emphasize that expressions \eqref{eq:18}, \eqref{eq:19} for polarization
of final electron obtained for pure initial state holds for mixed state
too.

Note also that for electron scattering in external field the initial pure spin
state leads to definite polarization of final state. It is not evident in
advance, quite the contrary, it would be natural to expect the appearance of a
mixed final state in this case but calculations of amplitude square demonstrates
this.

Both these facts are consequence of the equivalence of two problems:
$s^{(f)}=z$, discussed above.


\section{Spectral representation of propagator}
\label{sec:prop-spectral}

We want to apply the problem of looking for complete polarization axis
\eqref{eq:28} to the case of fermion in an intermediate state. All aforesaid is
easily extended for the case of virtual fermions if to use a spectral
representation of propagator.

In order to construct this representation one needs to solve eigenvalue problem
for inverse propagator $S(p)=G^{-1}(p)$. Since there is convenient
$\gamma$-matrix basis, one can solve a matrix problem: to look for
eigenprojectors $\Pi$ and eigenvalues $\lambda$
\begin{equation}\label{eq:5}
  S\Pi=\lambda\Pi.
\end{equation}
Having solved this problem, i.e. having found eigenvalues $\lambda_{i}$ and
orthogonal system of eigenprojectors $\Pi_{i}$
\begin{equation}\label{eq:57}
  \Pi_{i} \Pi_{k} = \delta_{ik} \Pi_{i} ,\quad
  i,k=1,2,
\end{equation}
we can construct the spectral representation of inverse propagator
\begin{equation}\label{eq:7}
  S(p)=\lambda_{1}\Pi_{1}+\lambda_{2}\Pi_{2}. 
\end{equation}
If the system of projectors is complete, then this expression can be easily
reversed and propagator looks like this:
\begin{equation}\label{eq:15}
  G(p)=\frac{1}{\lambda_{1}}\Pi_{1}+\frac{1}{\lambda_{2}}\Pi_{2} ,
\end{equation}
i.e. propagator poles are zeroes of eigenvalues $\lambda_{i}$. Let us consider
how the propagator spectral representation is shown up in particular cases.

The eigenprojectors $\Pi_{i}$ for a bare propagator are the known off-shell
projector operators $\Lambda^{\pm}_{W}$
\begin{equation}\label{eq:17}
  \Lambda^{\pm}_{W}=\frac{1}{2}\Big( 1\pm \frac{\hat{p}}{W} \Big),
  \quad p^{2}=W^{2},
\end{equation}
where $W$ is center-of-mass energy. As a result the bare propagator
\begin{equation}\label{eq:75}
  \begin{gathered}
    G_{0}(p)=\frac{1}{\hat{p}-m_{0}}=\frac{1}{W-m_{0}}\Lambda^{+}_{W}+
    \frac{1}{-W-m_{0}}\Lambda^{-}_{W},\\
    \lambda_{1}=W-m_{0},\quad
    \lambda_{2}=-W-m_{0},
  \end{gathered}
\end{equation}
is shown up as a sum of poles with positive and negative energies. It is
necessary to stress that we have covariant separation of poles $1/(W\pm m_{0})$
because $W$ is invariant mass. Besides that, since fermion and anti-fermion have
opposite parities the representation \eqref{eq:75} sorts out contributions by
parity\footnote{In contrast to equivalent electrons method \cite{Baier:1973ms,
    Baier:1980kx} (see also the usage of this method in concrete calculations
  \cite{Ginzburg:1994yu,Chen:1975sh,Fadin:1997sn}) the representation
  \eqref{eq:75} is based on the orthogonal projectors and has covariant
  form. Besides, it is easily generalized for the case of dressed propagator.}.

With account of interaction the inverse propagator takes the form
\begin{equation}
  \label{eq:23}
  S(p)=\hat{p}-m_{0}-\Sigma(p),
\end{equation}
where $\Sigma(p)$ is self-energy contribution. For dressed propagator the
spectral representation looks differently depending on interaction.
\begin{itemize}
\item If theory does not have $\gamma^{5}$ in a vertex, then $\Sigma(p)$
  contains unit matrix and $\hat{p}$
  \begin{equation}\label{eq:58}
    \Sigma(p)= A(p^{2})+\hat{p}
    B(p^{2})=\Sigma^{+}(W)\Lambda^{+}_{W}+\Sigma^{-}(W)\Lambda^{-}_{W},
  \end{equation}
  where $\Sigma^{\pm}(W)=A(W^{2}) \pm W B(W^{2})$. In this case eigenprojectors
  coincide with the such for bare propagator \eqref{eq:17} and the propagator
  spectral representation has form
  \begin{equation}\label{eq:35}
    G(p)=\frac{1}{\hat{p}-m_{0}-\Sigma(p)}=\frac{1}{W-m_{0}-\Sigma^{+}(W)}
    \Lambda^{+}_{W}+\frac{1}{-W-m_{0}-\Sigma^{-}(W)}\Lambda^{-}_{W}.
  \end{equation}
\item In theory with $\gamma^{5}$ the self-energy contribution also has
  $\gamma^{5}$ terms
  \begin{equation}\label{eq:41}
    \Sigma(p)=A(p^{2})+\hat{p}B(p^{2})+\gamma^{5}C(p^{2})+
    \hat{p}\gamma^{5}D(p^{2}),
  \end{equation}
  and eigenprojectors $\Pi_{i}$ do not coincide with $\Lambda^{\pm}_{W}$. In
  this case the eigenprojectors take a more complicated form
  \cite{Kaloshin:2011gy}:
  \begin{equation}\label{eq:43}
    \begin{gathered}
      \Pi_{1,2}(p) = \frac{1}{2}\big( 1\pm\hat{n}\tau \big),\quad
      \hat{n}=\frac{\hat{p}}{W}, \\
      \tau = \frac{1}{R}\Big( 1-B -\gamma^{5}D-\hat{n}\gamma^{5}\frac{C}{W}
      \Big), \quad R=\sqrt{(1-B)^{2}-D^{2}+C^{2}/W^{2}},
    \end{gathered}
  \end{equation}
  and eigenvalues $\lambda_{i}(W)$ are
  \begin{equation}\label{eq:44}
    \lambda_{1,2}(W)=-m_{0} - A(W^{2})\pm WR(W^{2}).
  \end{equation} 
\end{itemize}

An essential aspect related with the completeness of eigenprojectors system is
the existence of spin projectors commuting with propagator. The existence of
such projectors follows simply from counting of degrees of freedom in the
problem \eqref{eq:5}. However, the standard spin projectors
\begin{equation}\label{eq:38}
  \Sigma_{0}(s)=\frac{1+\gamma^{5}\hat{s}}{2},\quad
  s^{2}=-1,\quad
  (sp)=0,
\end{equation}
cease to commute with propagator in the presence of $\gamma^{5}$ in a vertex.

Nevertheless, there exist the generalized (dressed) spin projectors \cite{Kaloshin:2015rva}
closely related with eigenprojectors \eqref{eq:43} and having all desired
properties
\begin{equation}\label{eq:45}
  \Sigma(s)=\frac{1}{2}\big( 1+\gamma^{5}\hat{s}\tau \big).
\end{equation}

Note that $\Pi_{i}(p)$ and $\Sigma(s)$ have the same matrix factor $\tau$
\eqref{eq:43} which possesses the following property. Without interaction
($B=C=D=0$) or in theory with parity conservation ($C=D=0$) this factor turns
into unit matrix $\tau=1$. As a consequence the projectors $\Pi_{i}(p)$ and
$\Sigma(s)$ return to the standard (bare) form.

Furthermore, it can be seen that ``under the observation'' of energy
eigenprojector $\Pi_{i}(p)$ the spin projector \eqref{eq:45} is significantly
simplified
\begin{equation}\label{eq:45x}
  \Pi_{i}(p) \Sigma(s) = \Pi_{i}(p) \, \frac{1}{2}
  \big( 1+\gamma^{5}\hat{s}\hat{n} \big) .
\end{equation}

The problem of looking for complete polarization axis of bispinor \eqref{eq:28}
is naturally generalized for the case of virtual fermion.

Using free propagator \eqref{eq:17} one needs only to change projector
$\Lambda^{+}_{2}$ in \eqref{eq:28} to one of off-shell projectors
$\Lambda_{W}^{\pm} (p_{2})=(1\pm \hat{p}_{2}/W)/2$, $p_{2}^{2}=W^{2}$. The
problem of looking for complete polarization axis \eqref{eq:28} is turned into
that:
\begin{equation}\label{axe_off}
  \frac{1}{2}(1+\gamma^{5}\hat{z}^{\pm}) \cdot \Lambda_{W}^{\pm} (p_{2})\Gamma
  \Lambda^{+}_{1}\Sigma_{1}u_{1} = \Lambda_{W}^{\pm} (p_{2}) \Gamma
  \Lambda^{+}_{1}\Sigma_{1}u_{1},\quad
  (z^{\pm}p_{2})=0,
\end{equation}
and (see Appendix~\ref{sec:bispinor-polar-axis-existence}) such problem also has
solution: for any bispinor $u_{1}$ there exists a vector $z_{\mu}^{\pm}$,
$(z^{\pm})^{2}=-1$.

The aforesaid is also true for dressed propagator in theory with $\gamma^{5}$ as
well. Let us write down the problem of looking for complete polarization axis
for dressed energy and spin projectors \eqref{eq:43}, \eqref{eq:45}:
\begin{equation}\label{axe_dr}
  \Sigma({z}^{\pm}) \cdot \Pi^{\pm} (p_{2})\Gamma
  \Lambda^{+}_{1}\Sigma_{1}u_{1} = \Pi^{\pm} (p_{2}) \Gamma
  \Lambda^{+}_{1}\Sigma_{1}u_{1},\quad
  (z^{\pm}p_{2})=0.
\end{equation}
The dressed energy projectors \eqref{eq:43} in the case of $\mathsf{CP}$
conservation can be rewritten in form (it may be considered as $\gamma$-matrices
transformation $\gamma_{\mu}\to\gamma_{\mu}'$, see details in
Appendix~\ref{sec:gamma})
\begin{equation}
  \Pi^{\pm} =  \frac{1}{2} \big( 1 \pm \hat{n} \tau \big) =
  \frac{1}{2} \big( 1 \pm \hat{n} \exp^{2\kappa \gamma^{5}} \big) =
  \exp^{-\kappa \gamma^{5}} \frac{1}{2} \big( 1 \pm \hat{n}  \big)
  \exp^{\kappa \gamma^{5}}.
\end{equation}
The dressed spin projector can be presented by the same way
\begin{equation}
  \Sigma({z}) = \frac{1}{2} \big( 1 + \gamma^{5}\hat{z} \tau \big) =
  \exp^{-\kappa \gamma^{5}} \frac{1}{2} \big( 1 + \gamma^{5}\hat{z}  \big)
  \exp^{\kappa \gamma^{5}}.
\end{equation}
Thereafter the problem of looking for axis \eqref{axe_dr} takes form:
\begin{equation}\label{axe_2}
  \frac{1}{2} \big( 1 + \gamma^{5}\hat{z}^{\pm} \big) \cdot
  \Big( \Lambda_{W}^{\pm} (p_{2}) \exp^{\kappa \gamma^{5}} \Gamma
  \Lambda^{+}_{1}\Sigma_{1}u_{1} \Big) =
  \Big( \Lambda_{W}^{\pm} (p_{2}) \exp^{\kappa \gamma^{5}} \Gamma
  \Lambda^{+}_{1}\Sigma_{1}u_{1} \Big),\quad
  (z^{\pm}p_{2})=0.
\end{equation}
So the problem \eqref{axe_dr} for dressed energy projectors $\Pi^{\pm}(p_{2})$
is reduced to the problem \eqref{axe_off}, involving the bare off-shell
projectors $\Lambda^{\pm}_{W}(p_{2})$ \eqref{eq:17}. As a result one can
conclude that for the problem \eqref{axe_dr} also exists the axis of complete
polarization $z^{\pm}$.


\section{Polarization of fermion in an intermediate state}
\label{sec:intermed-fermion-polariz}

The spectral representation of propagator, where the orthogonal off-shell
projectors are arisen allows to give an accurate definition of fermion
polarization in an intermediate state. Thereafter the generalization of the
problem of looking for axis \eqref{eq:28} (replacing projectors by the off-shell
ones) yields the recipe of polarization calculation in an intermediate state.

Consider some process with intermediate fermion is born in $s$-channel. Here the
external boson lines correspond either to some on-mass-shell particles or to
external field.
\begin{center}
  \includegraphics{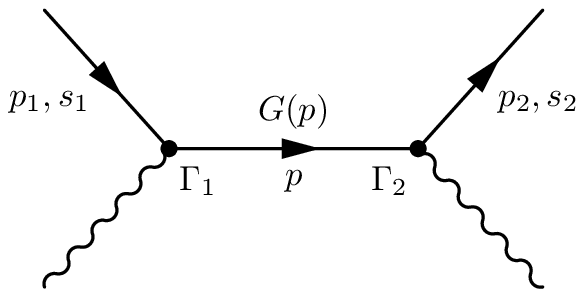}
\end{center}

The corresponding amplitude looks as
\begin{equation}\label{ampl}
  \mathscr{M}=\bar{u}_{2}(p_{2},s_{2})\Gamma_{1} G(p) \Gamma_{2}
  u_{1}(p_{1},s_{1}).
\end{equation}

For the case of bare propagator or theory without $\gamma^{5}$ the fermion
propagator in the intermediate state has form
\begin{equation}\label{eq:46}
  G(p)=\frac{1}{\lambda_{1}}\Lambda^{+}_{W}+\frac{1}{\lambda_{2}}\Lambda^{-}_{W},
\end{equation}
where $\Lambda^{\pm}_{W}$ are off-shell energy projectors \eqref{eq:17}.

If to recall the problem of looking for complete polarization axis involving
$\Lambda^{\pm}_{W}$ \eqref{axe_off}, the propagator in the amplitude
\eqref{ampl} can be rewritten as following
\begin{equation}\label{eq:47}
  G  = \frac{1}{\lambda_{1}}\Lambda^{+}_{W}\Sigma_{0}(z^{+}) +
  \frac{1}{\lambda_{2}}\Lambda^{-}_{W} \Sigma_{0}(z^{-}).
\end{equation}

This gives a natural definition for polarization of fermion in an intermediate
state. Polarization vectors $z^{\pm}$ are different for poles with positive and
negative energies. It is not so evident that spin density matrices in
\eqref{eq:47} are pure ones: $\big(z^{\pm}\big)^{2} = -1$, however, it follows
from the problem of looking for axis \eqref{axe_off}.

It is possible to calculate polarization vector either with the help of the
problem of looking for axis \eqref{axe_off} or by analogy with \eqref{eq:11}
with the use of spur:
\begin{equation}
  \Sp \big(\Lambda^{\pm}_{W} \Sigma_{0}(z^{\pm}) \Gamma u_{1}\bar{u}_{1}
  \tilde{\Gamma}\big)=A\Big(1+\frac{B_{\mu}}{A}z^{\pm \mu}
  \Big),\quad
  z^{\pm}_{\mu}=-\frac{B^{\perp}_{\mu}}{A}.
\end{equation}
It seems that the second way is technically simpler. As indicated above the
formula for polarization $z^{\pm}$ calculated for pure state holds for mixed
initial state also.

Let us take a look now on the case of the theory with $\gamma^{5}$ in vertex. In
this case dressed fermion propagator in intermediate state may be represented as
\begin{equation}\label{eq:54}
  G(p)=\frac{1}{\lambda_{1}}\Pi_{1}(p)+\frac{1}{\lambda_{2}}\Pi_{2}(p),
\end{equation}
where $\Pi_{1,2}(p)$ are the energy projectors \eqref{eq:43}. Using the problem
\eqref{axe_dr} one can see that the dressed propagator inside the diagram
acquires spin projectors
\begin{equation}\label{eq:56}
  G\to\tilde{G}=\frac{1}{\lambda_{1}}\Pi_{1}(p)\Sigma(z^{+})+
  \frac{1}{\lambda_{2}}\Pi_{2}(p)\Sigma(z^{-}), \quad (z^{\pm})^{2} = -1.
\end{equation}
It should be pointed out that dressed energy projectors $\Pi_{i}(p)$ presented
here contain self-energy contributions and should be renormalized.


\section{Conclusions}
\label{sec:conclusions}

The main statement of the work is that the final fermion polarization as such
\eqref{eq:14} can be calculated by a non-standard way, namely by using the
problem of search for complete polarization axis of bispinor \eqref{eq:27}. We
have shown that these two problems are equivalent.

It is unlikely that the problem \eqref{eq:27} leads to more simple calculations,
rather opposite, the calculation of spurs in the standard approach is a routine
operation which is successfully performed by a computer. However, in the problem
of looking for axis \eqref{eq:27} the amplitude is used instead of its
square. This allows to apply this problem to find polarization of fermion in an
intermediate state.

The essential moment is the use of propagator spectral representation
\eqref{eq:75}, \eqref{eq:35}, \eqref{eq:43}. This approach leads to the system
of orthogonal off-shell projectors and allows to give clear definition for
fermion polarization in an intermediate state. Thus it is possible to use either
free propagator or dressed one, including the theory with $\gamma^{5}$ in a
vertex.

When a propagator is written in form of the spectral representation it is
possible to use the problem of looking for axis to calculate the off-mass-shell
polarization. Indeed, in the problem \eqref{eq:27} of final electron
polarization (on-mass-shell) the energy projector is changed to the off-shell
one. As a result the propagator inside of a diagram takes view \eqref{eq:56}
where spin density matrices correspond to complete polarization.

The fact that, for example, the projector $\Pi_{1}$ of positive energy state in
\eqref{eq:56} is accompanied by completely polarized spin density matrix
$\Sigma(z^{+})$, $(z^{+})^{2}=-1$ is not evident in advance and is the result of
the problem of looking for axis.

The evident application of suggested approach is concerned with $t$-quark
polarization but it needs single consideration.

We are grateful to V.G. Serbo for fruitful discussions and participation in the
beginning of the work.

\appendix


\section{Auxiliary matrix problem}
\label{sec:matrix-prob}

We want to find the most general form of matrix $X$ obeying equations
\begin{equation}\label{eq:6}
  X(1-\hat{n})=(1-\hat{n})X=0,\quad
  n^{\mu}=p^{\mu}/m,\quad
  n^{2}=1.
\end{equation}
For this purpose we decompose the matrix $X$
\begin{equation}\label{eq:42}
  X=x_{0}+x_{\mu}\gamma^{\mu}+\bar{x}_{0}\gamma^{5}+\bar{x}_{\mu}\gamma^{5}
  \gamma^{\mu}+x_{\mu\nu}\sigma^{\mu\nu},
\end{equation}
and determine its coefficients.

Computing coefficients at $\gamma^{5}$ in relations \eqref{eq:6} one obtain pair
of equations
\begin{equation}
  \label{eq:32}
  \bar{x}_{0}+(n \bar{x})=0,\quad
  \bar{x}_{0}-(n\bar{x})=0,
\end{equation}
from which it immediately follows that $\bar{x}_{0}=0$ and $(n \bar{x})=0$.

Next, determining coefficients at $\gamma^{5}\gamma^{\mu}$ in these relations we
come to equations
\begin{equation}
  \label{eq:50}
  x_{\mu}-x_{0}n_{\mu}+2x_{\mu\nu}n^{\nu}=0,\quad
  x_{\mu}-x_{0}n_{\mu}-2x_{\mu\nu}n^{\nu}=0.
\end{equation}
Hence it follows that $x_{\mu}=x_{0}n_{\mu}$.

Finally, calculating coefficients at $\sigma^{\mu\nu}$ in any of relations
\eqref{eq:6} we determine $x_{\mu\nu}$:
\begin{equation}
  \label{eq:51}
  x_{\mu\nu}=-\frac{\imath}{2}\epsilon_{\mu\nu\alpha\beta}n^{\alpha}
  \bar{x}^{\beta}.
\end{equation}
Thus the matrix $X$ has view
\begin{equation}\label{eq:77}
  X=x_{0}(1+\hat{n})+\bar{x}_{\mu}\gamma^{5}\gamma^{\mu}-\frac{\imath}{2}
  \epsilon_{\mu\nu\alpha\beta}n^{\alpha}\bar{x}^{\beta}\sigma^{\mu\nu},
\end{equation}
with arbitrary parameter $x_{0}$ and vector $\bar{x}_{\mu}$: $(n\bar{x})=0$. Let
us rewrite this expression in more compact form
\begin{equation}
  \label{eq:52}
  X=x_{0}(1+\hat{n})+\bar{x}_{\mu}\gamma^{5}\gamma^{\mu}(1+\hat{n})
  =(1+\hat{n})(x_{0}+\bar{x}_{\mu}\gamma^{5}\gamma^{\mu}),\quad
  (n\bar{x})=0.
\end{equation}


\section{Complete polarization axis of bispinor}
\label{sec:bispinor-polar-axis-existence}

It is known that any spinor has complete polarization axis. Bispinor also has
similar property: if $\Psi(p)$ is an arbitrary solution of Dirac equation then
there exists such vector $s_{\mu}$ that (cp. with \eqref{eq:3})
\begin{equation}\label{eq:20} 
  \gamma^{5}\hat{s}\cdot\Psi(p)=\Psi(p),\quad
  s^{2}=-1,\quad (sp)=0,\quad p^{2}=m^{2}.
\end{equation}

Let us show this for solution of Dirac equation with positive energy
$u(p,s)$. Write down equation \eqref{eq:20} in split form
\begin{equation}\label{eq:24}
  \gamma^{5}\hat{s}
  \begin{pmatrix}
    u_{1}\\
    u_{2}
  \end{pmatrix}=
  \begin{pmatrix}
    u_{1}\\
    u_{2}
  \end{pmatrix}.
\end{equation}
In the standard representation of the $\gamma$-matrices we have
\begin{equation*}
  \gamma^{5}\hat{s}=
  \begin{pmatrix}
    \vect{\sigma}\cdot\vect{s} & -s_{0}\\
    s_{0} & -\vect{\sigma}\cdot\vect{s}
  \end{pmatrix},\quad
  u=
  \begin{pmatrix}
    (n_{0}+1)\phi\\
    \vect{n}\cdot\vect{\sigma}\phi
  \end{pmatrix},
\end{equation*}
where $n^{\mu}=p^{\mu}/m$, $n^{2}=1$, $n^{0} > 0$ and $\phi$ is an arbitrary
two-component spinor. Substituting these formulas into \eqref{eq:24} we come to
the problem of determination of complete polarization axis for two-dimensional
spinor
\begin{equation}\label{eq:25}
  \vect{\sigma}\cdot \left( \vect{s}-\frac{s_{0}\vect{n}}{n_{0}+1} \right)
  \phi=\phi.
\end{equation}
It is easy to check that vector appeared here
\begin{equation}
  \vect{S}=\vect{s}-\frac{s_{0}\vect{n}}{n_{0}+1}
\end{equation}
has unit length $\vect{S}^{2}=1$ if to recall that $s^{2}=-1$, $(ns)=0$ and
$n^{2}=1$.

One can conclude that for solutions of Dirac equation with positive energy the
relation \eqref{eq:20} can be reduced to equation \eqref{eq:25} which has
solution for any two-dimensional column (cp. with \eqref{eq:3}). Of course all
the same holds for solutions with negative energy.

The off-shell bispinors are defined as solutions of equations
$\Lambda_{W}^{-}\Psi(p)=0$ and $\Lambda_{W}^{+}\Psi(p)=0$ at $p^{2}=W^{2}$
(positive and negative frequency solutions). It is easy to check that there
exists complete polarization axis $s_{\mu}$ for such bispinors, in other words
\begin{equation}
  \label{eq:60}
  \gamma^{5}\hat{s}\Psi(p)=\Psi(p),\quad
  s^{2}=-1,\quad
  (ps)=0,\quad
  p^{2}=W^{2}.
\end{equation}
The proof repeats exactly the similar consideration for Dirac equation solution.


\section{Dressed energy and spin projectors}
\label{sec:gamma}

Let us show that dressed energy projector \eqref{eq:43} (in the theory with
$\gamma^{5}$) can be transformed into the standard one \eqref{eq:17} by
transition to another representation of $\gamma$-matrices. At first, let us
consider the case of $\mathsf{CP}$ conservation ($C=0$ in \eqref{eq:43}) and
show that in such case
\begin{equation}
  \Pi_{i} = \frac{1}{2}\big( 1+\hat{n}\tau \big) =
  \frac{1}{2} \left( 1+\hat{n} \frac{(1-B)-D\gamma^{5}}
    {\sqrt{(1-B)^{2} - D^{2}}}\right) =
  \frac{1}{2}\big( 1+n_{\mu} \gamma^{\prime\mu} \big) .
\end{equation}
with some transformed matrices $\gamma^{\prime\mu}$.

Firstly, note that matrix $\tau$ can be written in exponential view
\begin{equation}\label{eq:71}
  \tau=\frac{(1-B)-D\gamma^{5}}{\sqrt{(1-B)^{2} - D^{2}}} =
  \cosh{2\kappa}+\gamma^{5}\sinh{2\kappa}=\exp^{2\kappa\gamma^{5}}.
\end{equation}
Next, performing transformation of form
\begin{equation*}
  \hat{n}\tau = \hat{n}\exp^{2\kappa\gamma^{5}}=
  \hat{n}\exp^{\kappa\gamma^{5}} \cdot \exp^{\kappa\gamma^{5}} =
  \exp^{-\kappa\gamma^{5}} \hat{n}  \exp^{\kappa\gamma^{5}} =
  n^{\mu} \gamma^{\prime}_{\mu},
\end{equation*}
we see that $\gamma$-matrices in another representation are appeared
\begin{equation}\label{eq:72}  
  \gamma_{\mu}'=T^{-1} \gamma_{\mu}T,\quad
  T=\exp^{\kappa\gamma^{5}},\quad
  \gamma^{\prime 5}=\gamma^{5}.
\end{equation}

In the general case when $\mathsf{CP}$ is not conserved the $\tau$ can contain
matrix $\hat{n}\gamma^{5}$ as well. In that case in order to reduce projector
into standard view a one more transformation of $\gamma$-matrices is required
\begin{equation}\label{eq:73}
  \gamma_{\mu} \to \gamma_{\mu}^{\prime} \to \gamma_{\mu}^{\prime\prime},\quad\quad
  \gamma_{\mu}^{\prime\prime} = \exp^{-\beta\hat{n}^{\prime}\gamma^{\prime 5}}
  \gamma_{\mu}^{\prime} \exp^{\beta\hat{n}^{\prime}\gamma^{\prime 5}}.
\end{equation}
Finally, let us note that transformations \eqref{eq:72}, \eqref{eq:73} are not
unitary.



\begin{thebibliography}{17}%
\makeatletter
\providecommand \@ifxundefined [1]{%
 \@ifx{#1\undefined}
}%
\providecommand \@ifnum [1]{%
 \ifnum #1\expandafter \@firstoftwo
 \else \expandafter \@secondoftwo
 \fi
}%
\providecommand \@ifx [1]{%
 \ifx #1\expandafter \@firstoftwo
 \else \expandafter \@secondoftwo
 \fi
}%
\providecommand \natexlab [1]{#1}%
\providecommand \enquote  [1]{``#1''}%
\providecommand \bibnamefont  [1]{#1}%
\providecommand \bibfnamefont [1]{#1}%
\providecommand \citenamefont [1]{#1}%
\providecommand \href@noop [0]{\@secondoftwo}%
\providecommand \href [0]{\begingroup \@sanitize@url \@href}%
\providecommand \@href[1]{\@@startlink{#1}\@@href}%
\providecommand \@@href[1]{\endgroup#1\@@endlink}%
\providecommand \@sanitize@url [0]{\catcode `\\12\catcode `\$12\catcode
  `\&12\catcode `\#12\catcode `\^12\catcode `\_12\catcode `\%12\relax}%
\providecommand \@@startlink[1]{}%
\providecommand \@@endlink[0]{}%
\providecommand \url  [0]{\begingroup\@sanitize@url \@url }%
\providecommand \@url [1]{\endgroup\@href {#1}{\urlprefix }}%
\providecommand \urlprefix  [0]{URL }%
\providecommand \Eprint [0]{\href }%
\providecommand \doibase [0]{http://dx.doi.org/}%
\providecommand \selectlanguage [0]{\@gobble}%
\providecommand \bibinfo  [0]{\@secondoftwo}%
\providecommand \bibfield  [0]{\@secondoftwo}%
\providecommand \translation [1]{[#1]}%
\providecommand \BibitemOpen [0]{}%
\providecommand \bibitemStop [0]{}%
\providecommand \bibitemNoStop [0]{.\EOS\space}%
\providecommand \EOS [0]{\spacefactor3000\relax}%
\providecommand \BibitemShut  [1]{\csname bibitem#1\endcsname}%
\let\auto@bib@innerbib\@empty
\bibitem [{\citenamefont {Berestetskii}\ \emph {et~al.}(2012)\citenamefont
  {Berestetskii}, \citenamefont {Pitaevskii},\ and\ \citenamefont
  {Lifshitz}}]{Berestetskii:2012quantum}%
  \BibitemOpen
  \bibfield  {author} {\bibinfo {author} {\bibfnamefont {V.}~\bibnamefont
  {Berestetskii}}, \bibinfo {author} {\bibfnamefont {L.}~\bibnamefont
  {Pitaevskii}}, \ and\ \bibinfo {author} {\bibfnamefont {E.}~\bibnamefont
  {Lifshitz}},\ }\href@noop {} {\emph {\bibinfo {title} {Quantum
  Electrodynamics}}},\ Vol.~\bibinfo {volume} {4}\ (\bibinfo  {publisher}
  {Elsevier Science},\ \bibinfo {year} {2012})\BibitemShut {NoStop}%
\bibitem [{\citenamefont {Budnev}\ \emph {et~al.}(1975)\citenamefont {Budnev},
  \citenamefont {Ginzburg}, \citenamefont {Meledin},\ and\ \citenamefont
  {Serbo}}]{Budnev:1974de}%
  \BibitemOpen
  \bibfield  {author} {\bibinfo {author} {\bibfnamefont {V.~M.}\ \bibnamefont
  {Budnev}}, \bibinfo {author} {\bibfnamefont {I.~F.}\ \bibnamefont
  {Ginzburg}}, \bibinfo {author} {\bibfnamefont {G.~V.}\ \bibnamefont
  {Meledin}}, \ and\ \bibinfo {author} {\bibfnamefont {V.~G.}\ \bibnamefont
  {Serbo}},\ }\href {\doibase 10.1016/0370-1573(75)90009-5} {\bibfield
  {journal} {\bibinfo  {journal} {Phys. Rept.}\ }\textbf {\bibinfo {volume}
  {15}},\ \bibinfo {pages} {181} (\bibinfo {year} {1975})}\BibitemShut
  {NoStop}%
\bibitem [{\citenamefont {Aaltonen}\ \emph {et~al.}(2011)\citenamefont
  {Aaltonen} \emph {et~al.}}]{Aaltonen:2010nz}%
  \BibitemOpen
  \bibfield  {author} {\bibinfo {author} {\bibfnamefont {T.}~\bibnamefont
  {Aaltonen}} \emph {et~al.} (\bibinfo {collaboration} {CDF}),\ }\href
  {\doibase 10.1103/PhysRevD.83.031104} {\bibfield  {journal} {\bibinfo
  {journal} {Phys. Rev.}\ }\textbf {\bibinfo {volume} {D83}},\ \bibinfo {pages}
  {031104} (\bibinfo {year} {2011})},\ \Eprint {http://arxiv.org/abs/1012.3093}
  {arXiv:1012.3093 [hep-ex]} \BibitemShut {NoStop}%
\bibitem [{\citenamefont {Abazov}\ \emph {et~al.}(2012)\citenamefont {Abazov}
  \emph {et~al.}}]{Abazov:2011gi}%
  \BibitemOpen
  \bibfield  {author} {\bibinfo {author} {\bibfnamefont {V.~M.}\ \bibnamefont
  {Abazov}} \emph {et~al.} (\bibinfo {collaboration} {D0}),\ }\href {\doibase
  10.1103/PhysRevLett.108.032004} {\bibfield  {journal} {\bibinfo  {journal}
  {Phys. Rev. Lett.}\ }\textbf {\bibinfo {volume} {108}},\ \bibinfo {pages}
  {032004} (\bibinfo {year} {2012})},\ \Eprint {http://arxiv.org/abs/1110.4194}
  {arXiv:1110.4194 [hep-ex]} \BibitemShut {NoStop}%
\bibitem [{\citenamefont {Chatrchyan}\ \emph {et~al.}(2014)\citenamefont
  {Chatrchyan} \emph {et~al.}}]{Chatrchyan:2013wua}%
  \BibitemOpen
  \bibfield  {author} {\bibinfo {author} {\bibfnamefont {S.}~\bibnamefont
  {Chatrchyan}} \emph {et~al.} (\bibinfo {collaboration} {CMS}),\ }\href
  {\doibase 10.1103/PhysRevLett.112.182001} {\bibfield  {journal} {\bibinfo
  {journal} {Phys. Rev. Lett.}\ }\textbf {\bibinfo {volume} {112}},\ \bibinfo
  {pages} {182001} (\bibinfo {year} {2014})},\ \Eprint
  {http://arxiv.org/abs/1311.3924} {arXiv:1311.3924 [hep-ex]} \BibitemShut
  {NoStop}%
\bibitem [{\citenamefont {Aad}\ \emph {et~al.}(2013)\citenamefont {Aad} \emph
  {et~al.}}]{Aad:2013ksa}%
  \BibitemOpen
  \bibfield  {author} {\bibinfo {author} {\bibfnamefont {G.}~\bibnamefont
  {Aad}} \emph {et~al.} (\bibinfo {collaboration} {ATLAS}),\ }\href {\doibase
  10.1103/PhysRevLett.111.232002} {\bibfield  {journal} {\bibinfo  {journal}
  {Phys. Rev. Lett.}\ }\textbf {\bibinfo {volume} {111}},\ \bibinfo {pages}
  {232002} (\bibinfo {year} {2013})},\ \Eprint {http://arxiv.org/abs/1307.6511}
  {arXiv:1307.6511 [hep-ex]} \BibitemShut {NoStop}%
\bibitem [{\citenamefont {Schilling}(2012)}]{Schilling:2012dx}%
  \BibitemOpen
  \bibfield  {author} {\bibinfo {author} {\bibfnamefont {F.-P.}\ \bibnamefont
  {Schilling}},\ }\href {\doibase 10.1142/S0217751X12300165} {\bibfield
  {journal} {\bibinfo  {journal} {Int. J. Mod. Phys.}\ }\textbf {\bibinfo
  {volume} {A27}},\ \bibinfo {pages} {1230016} (\bibinfo {year} {2012})},\
  \Eprint {http://arxiv.org/abs/1206.4484} {arXiv:1206.4484 [hep-ex]}
  \BibitemShut {NoStop}%
\bibitem [{\citenamefont {Bernreuther}\ and\ \citenamefont
  {Uwer}(2015)}]{Bernreuther:2015wqa}%
  \BibitemOpen
  \bibfield  {author} {\bibinfo {author} {\bibfnamefont {W.}~\bibnamefont
  {Bernreuther}}\ and\ \bibinfo {author} {\bibfnamefont {P.}~\bibnamefont
  {Uwer}},\ }\bibfield  {booktitle} {\emph {\bibinfo {booktitle} {{Proceedings,
  Advances in Computational Particle Physics: Final Meeting (SFB-TR-9)}}},\
  }\href {\doibase 10.1016/j.nuclphysbps.2015.03.025} {\bibfield  {journal}
  {\bibinfo  {journal} {Nucl. Part. Phys. Proc.}\ }\textbf {\bibinfo {volume}
  {261-262}},\ \bibinfo {pages} {414} (\bibinfo {year} {2015})}\BibitemShut
  {NoStop}%
\bibitem [{\citenamefont {Bjorken}\ and\ \citenamefont
  {Drell}(1964)}]{Bjorken:1964zz}%
  \BibitemOpen
  \bibfield  {author} {\bibinfo {author} {\bibfnamefont {J.~D.}\ \bibnamefont
  {Bjorken}}\ and\ \bibinfo {author} {\bibfnamefont {S.~D.}\ \bibnamefont
  {Drell}},\ }\href@noop {} {\emph {\bibinfo {title} {{Relativistic quantum
  mechanics}}}}\ (\bibinfo  {publisher} {McGraw-Hill},\ \bibinfo {year}
  {1964})\BibitemShut {NoStop}%
\bibitem [{\citenamefont {Baier}\ \emph {et~al.}(1973)\citenamefont {Baier},
  \citenamefont {Fadin},\ and\ \citenamefont {Khoze}}]{Baier:1973ms}%
  \BibitemOpen
  \bibfield  {author} {\bibinfo {author} {\bibfnamefont {V.}~\bibnamefont
  {Baier}}, \bibinfo {author} {\bibfnamefont {V.~S.}\ \bibnamefont {Fadin}}, \
  and\ \bibinfo {author} {\bibfnamefont {V.~A.}\ \bibnamefont {Khoze}},\ }\href
  {\doibase 10.1016/0550-3213(73)90291-5} {\bibfield  {journal} {\bibinfo
  {journal} {Nucl.Phys.}\ }\textbf {\bibinfo {volume} {B65}},\ \bibinfo {pages}
  {381} (\bibinfo {year} {1973})}\BibitemShut {NoStop}%
\bibitem [{\citenamefont {Baier}\ \emph {et~al.}(1981)\citenamefont {Baier},
  \citenamefont {Kuraev}, \citenamefont {Fadin},\ and\ \citenamefont
  {Khoze}}]{Baier:1980kx}%
  \BibitemOpen
  \bibfield  {author} {\bibinfo {author} {\bibfnamefont {V.~N.}\ \bibnamefont
  {Baier}}, \bibinfo {author} {\bibfnamefont {E.~A.}\ \bibnamefont {Kuraev}},
  \bibinfo {author} {\bibfnamefont {V.~S.}\ \bibnamefont {Fadin}}, \ and\
  \bibinfo {author} {\bibfnamefont {V.~A.}\ \bibnamefont {Khoze}},\ }\href
  {\doibase 10.1016/0370-1573(81)90140-X} {\bibfield  {journal} {\bibinfo
  {journal} {Phys. Rept.}\ }\textbf {\bibinfo {volume} {78}},\ \bibinfo {pages}
  {293} (\bibinfo {year} {1981})}\BibitemShut {NoStop}%
\bibitem [{\citenamefont {Kaloshin}\ and\ \citenamefont
  {Lomov}(2012)}]{Kaloshin:2011gy}%
  \BibitemOpen
  \bibfield  {author} {\bibinfo {author} {\bibfnamefont {A.}~\bibnamefont
  {Kaloshin}}\ and\ \bibinfo {author} {\bibfnamefont {V.}~\bibnamefont
  {Lomov}},\ }\href {\doibase 10.1140/epjc/s10052-012-2094-y} {\bibfield
  {journal} {\bibinfo  {journal} {Eur.Phys.J.}\ }\textbf {\bibinfo {volume}
  {C72}},\ \bibinfo {pages} {2094} (\bibinfo {year} {2012})},\ \Eprint
  {http://arxiv.org/abs/1111.1284} {arXiv:1111.1284 [hep-ph]} \BibitemShut
  {NoStop}%
\bibitem [{\citenamefont {Kaloshin}\ and\ \citenamefont
  {Lomov}(2016)}]{Kaloshin:2015rva}%
  \BibitemOpen
  \bibfield  {author} {\bibinfo {author} {\bibfnamefont {A.~E.}\ \bibnamefont
  {Kaloshin}}\ and\ \bibinfo {author} {\bibfnamefont {V.~P.}\ \bibnamefont
  {Lomov}},\ }\href {\doibase 10.1142/S0217751X16500317} {\bibfield  {journal}
  {\bibinfo  {journal} {Int. J. Mod. Phys.}\ }\textbf {\bibinfo {volume}
  {A31}},\ \bibinfo {pages} {1650031} (\bibinfo {year} {2016})},\ \Eprint
  {http://arxiv.org/abs/1501.06337} {arXiv:1501.06337 [hep-ph]} \BibitemShut
  {NoStop}%
\bibitem [{\citenamefont {Landau}\ and\ \citenamefont
  {Lifshitz}(2013)}]{Landau:2013quantum}%
  \BibitemOpen
  \bibfield  {author} {\bibinfo {author} {\bibfnamefont {L.}~\bibnamefont
  {Landau}}\ and\ \bibinfo {author} {\bibfnamefont {E.}~\bibnamefont
  {Lifshitz}},\ }\href@noop {} {\emph {\bibinfo {title} {Quantum Mechanics:
  Non-Relativistic Theory}}},\ \bibinfo {series} {Teoreticheskaia fizika},
  Vol.~\bibinfo {volume} {3}\ (\bibinfo  {publisher} {Elsevier Science},\
  \bibinfo {year} {2013})\BibitemShut {NoStop}%
\bibitem [{\citenamefont {Ginzburg}\ and\ \citenamefont
  {Serbo}(1994)}]{Ginzburg:1994yu}%
  \BibitemOpen
  \bibfield  {author} {\bibinfo {author} {\bibfnamefont {I.~F.}\ \bibnamefont
  {Ginzburg}}\ and\ \bibinfo {author} {\bibfnamefont {V.~G.}\ \bibnamefont
  {Serbo}},\ }\href {\doibase 10.1103/PhysRevD.49.2623} {\bibfield  {journal}
  {\bibinfo  {journal} {Phys. Rev.}\ }\textbf {\bibinfo {volume} {D49}},\
  \bibinfo {pages} {2623} (\bibinfo {year} {1994})}\BibitemShut {NoStop}%
\bibitem [{\citenamefont {Chen}\ and\ \citenamefont
  {Zerwas}(1975)}]{Chen:1975sh}%
  \BibitemOpen
  \bibfield  {author} {\bibinfo {author} {\bibfnamefont {M.-S.}\ \bibnamefont
  {Chen}}\ and\ \bibinfo {author} {\bibfnamefont {P.~M.}\ \bibnamefont
  {Zerwas}},\ }\href {\doibase 10.1103/PhysRevD.12.187} {\bibfield  {journal}
  {\bibinfo  {journal} {Phys. Rev.}\ }\textbf {\bibinfo {volume} {D12}},\
  \bibinfo {pages} {187} (\bibinfo {year} {1975})}\BibitemShut {NoStop}%
\bibitem [{\citenamefont {Fadin}\ \emph {et~al.}(1997)\citenamefont {Fadin},
  \citenamefont {Khoze},\ and\ \citenamefont {Martin}}]{Fadin:1997sn}%
  \BibitemOpen
  \bibfield  {author} {\bibinfo {author} {\bibfnamefont {V.~S.}\ \bibnamefont
  {Fadin}}, \bibinfo {author} {\bibfnamefont {V.~A.}\ \bibnamefont {Khoze}}, \
  and\ \bibinfo {author} {\bibfnamefont {A.~D.}\ \bibnamefont {Martin}},\
  }\href {\doibase 10.1103/PhysRevD.56.484} {\bibfield  {journal} {\bibinfo
  {journal} {Phys. Rev.}\ }\textbf {\bibinfo {volume} {D56}},\ \bibinfo {pages}
  {484} (\bibinfo {year} {1997})},\ \Eprint
  {http://arxiv.org/abs/hep-ph/9703402} {arXiv:hep-ph/9703402 [hep-ph]}
  \BibitemShut {NoStop}%
\end{thebibliography}
\providecommand{\particle}[1]{\mathrm{#1}}

\end{document}